\documentclass[twocolumn]{article}
\usepackage{arxiv}  %

\usepackage{physics} 
\usepackage{enumerate} 
\usepackage{pgfplots}
\pgfplotsset{compat=1.17} 
\usepackage{pgfplotstable}
\usepackage{tikz,pgfplots}
\usepackage{caption}
\usepackage{tipa}
\usepackage{float}
\usepackage{makecell}

\usepackage{subcaption}
\usepackage{multirow}
\usepackage[hang,flushmargin]{footmisc}

\newcommand{\ie}{\emph{i.e.,} }

\newcommand{\quotes}[1]{``#1''}

\usepackage[pdfencoding=auto, pdfusetitle]{hyperref}
\hypersetup{
  hidelinks,
  urlcolor=red,
  pdftitle={TITLE},
  pdfauthor={AUTHOR1, AUTHOR2, AUTHOR3},
  pdfkeywords={KEYWORD1, KEYWORD2, KEYWORD3},
}

\title{Preliminary Study on SSCF-derived Polar Coordinate for ASR}

\author{Sotheara Leang$^{1, 2}$, Eric Castelli$^2$, Dominique Vaufreydaz$^2$, Sethserey Sam$^1$\vspace{0.1cm}\\
{$^1$ Institute of Digital Research and Innovation, CADT, Phnom Penh, Cambodia}\\ %
{$^2$ Univ. Grenoble Alpes, CNRS, Grenoble INP, LIG, 38000 Grenoble, France}\\
}

\begin{document}

\twocolumn[{%
  \begin{@twocolumnfalse}
    \maketitle
  \end{@twocolumnfalse}
}]

\setcounter{footnote}{0}

\begin{abstract}

The transition angles are defined to describe the vowel-to-vowel transitions in the acoustic space of the Spectral Subband Centroids, and the findings show that they are similar among speakers and speaking rates. In this paper, we propose to investigate the usage of polar coordinates in favor of angles to describe a speech signal by characterizing its acoustic trajectory and using them in Automatic Speech Recognition. According to the experimental results evaluated on the BRAF100 dataset, the polar coordinates achieved significantly higher accuracy than the angles in the mixed and cross-gender speech recognitions, demonstrating that these representations are superior at defining the acoustic trajectory of the speech signal. Furthermore, the accuracy was significantly improved when they were utilized with their first and second-order derivatives ($\Delta$, $\Delta\Delta$), especially in cross-female recognition. However, the results showed they were not much more gender-independent than the conventional  Mel-frequency Cepstral Coefficients (MFCCs).

\end{abstract}

\section{Introduction}

Automatic Speech Recognition (ASR) plays a vital role in human-computer interactions, particularly in voice assistants. It allows such intelligent devices to translate a speech signal into textual information to obtain semantic comprehension before taking action. Due to the rapid growth of related disciplines, ASR systems have remarkably improved and are extensively used in several sectors, significantly improving work efficiency and reducing human demands.

Most state-of-the-art ASR systems based on Hidden Markov Model (HMM) or Deep Learning characterize the speech signal with acoustic features derived from the absolute frequency measurements such as Mel-frequency Cepstral Coefficients (MFCCs),  Perceptual Linear Predictive Cepstrum (PLP) and Mel-filter Bank \cite{saksamudre2015review, gupta2018state}. However, it is well known that the frequency space of speakers varies greatly, especially if we compare that of men with that of women or children: women and children produce speech with higher frequencies than men \cite{janse2014comparative, menard2002auditory, potamianos2003robust, hillenbrand2009role, carre2017speech}. Furthermore, if the physiological, emotional, or sociological factors (\ie age, speaking rate, accent) are considered, the acoustic space varies significantly due to the inter-speaker and intra-speaker differences \cite{benzeghiba2007automatic}. 

To take into account the large variability of speech (inter-speaker variability, environmental variability, etc.), current ASR systems require a very large amount of data for their training. This is even more true in the case of systems based on Deep Learning. This is a major handicap for developing such systems for poorly endowed or endangered languages. It is, in fact, often quite challenging and very costly to develop an ASR system for minority or low-resource languages when a large dataset of the transcribed speech does not exist \cite{besacier2014automatic, yu2020acoustic}. One research direction to reduce the training data size is to make the ASR systems intrinsically speaker-independent. If this is possible for the acoustic part of the systems, only one speaker would be sufficient for the learning phase.

This research proposes to describe the speech signal based on \quotes{acoustic gestures}, with the hypothesis that it is more robust and capable of improving the performance of ASR systems, particularly with respect to the natural variability of the speech. We suggest defining the acoustic gestures of the speech signal using polar coordinates rather than the transition angles on the acoustic space of the Spectral Subband Centroid Frequency (SSCF). We anticipate that the polar coordinates provide a better representation of the speech because they can define not only the transition direction but also the acoustic trajectory of the speech signal.

This paper is structured as follows: Section 2 includes related work. The proposed method is described in Section 3. The experiments are introduced in Section 4. Section 5 contains the results and discussions. Finally, section 6 presents the conclusion and future work.

\section{Related Work}

The Spectral Subband Centroids (SSCs) were examined and utilized as complement parameters to the cepstral features for speech recognition. They have properties similar to the formant frequencies and are robust to additive Gaussian noise \cite{paliwal1998spectral}. The SSC frequencies (SSCFs) were computed by dividing the frequency band into a number of subbands and computing the centroid of each subband using power spectrum of the speech signal (Equation \ref{formula_sscf}).

\begin{equation}\label{formula_sscf}
SSCF_m= \frac{\int_{l_m}^{h_m} fw_m(f)P^\gamma(f) \,df}{\int_{l_m}^{h_m} w_m(f)P^\gamma(f) \,df}
\end{equation}

Where ${l_m}$ and ${h_m}$ are the lower and upper bounds of subband ${m}$; ${w_m}$ is the subband filter; ${P(f)}$ is the power spectrum at frequency bin ${f}$; ${\gamma}$ is a coefficient regulating dynamic range of the power spectrum.

The SSCFs were subsequently explored in Vietnamese vowel-to-vowel (VV) transitions. Six subband filters were proposed to capture more information about the speech signal by splitting the frequency band in mel-scale into the equals-size subbands. The study found that the SSCF parameters have similar shape to the formant frequencies and provide a good estimation and continuous values, even during the consonant production \cite{tran2016acoustic}.

\begin{figure}[!htb]
\begin{center}
\includegraphics[width=7cm, height=6cm]{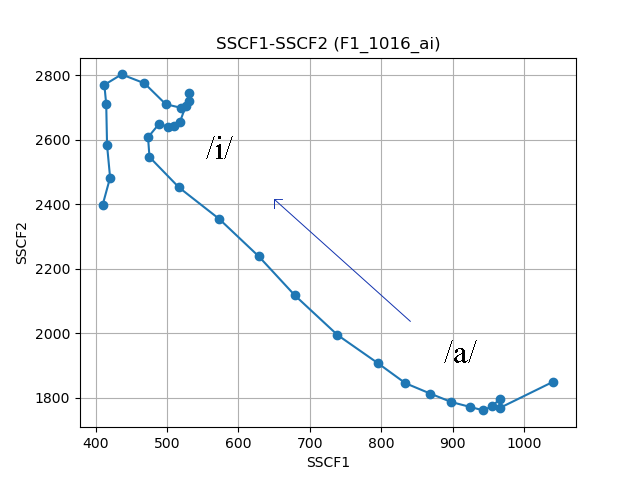}
\caption{The spectral transition of /ai/ on the SSCF1-SSCF2 plane produced by a Vietnamese female speaker}
\label{fig_ai_transition}
\end{center}
\end{figure}

\begin{figure}[!htb]
\center{\includegraphics[width=7.3cm, height=6cm]{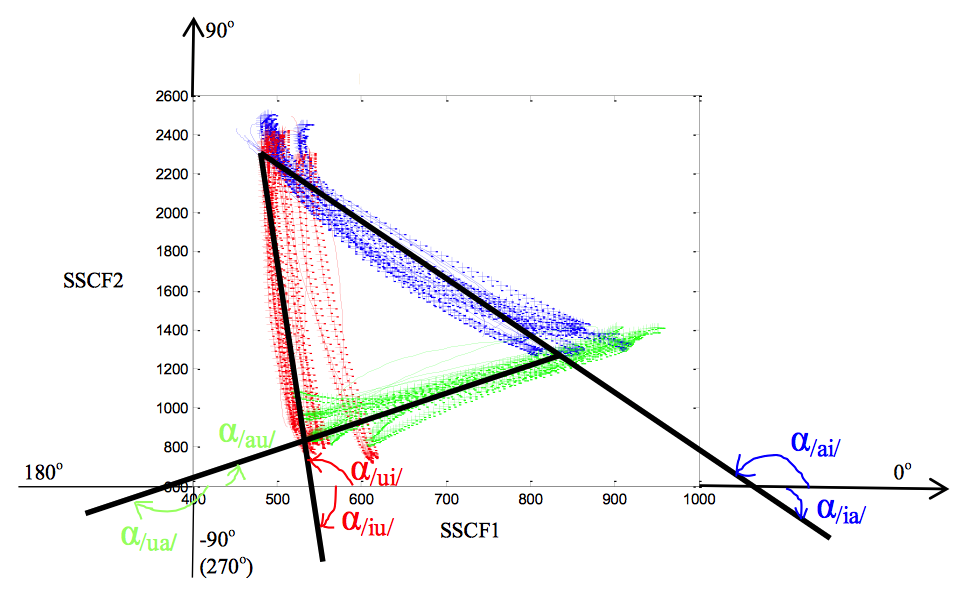}}
\caption{The SSCF angles of six vowel-to-vowel transitions on the SSCF1-SSCF2 plane produced by a Vietnamese male speaker (source \protect\cite{tran2016acoustic}).}
\label{fig_sscf_angle}
\end{figure}

\begin{figure}[htbp]
\center{\includegraphics[width=7.5cm, height=5.5cm]{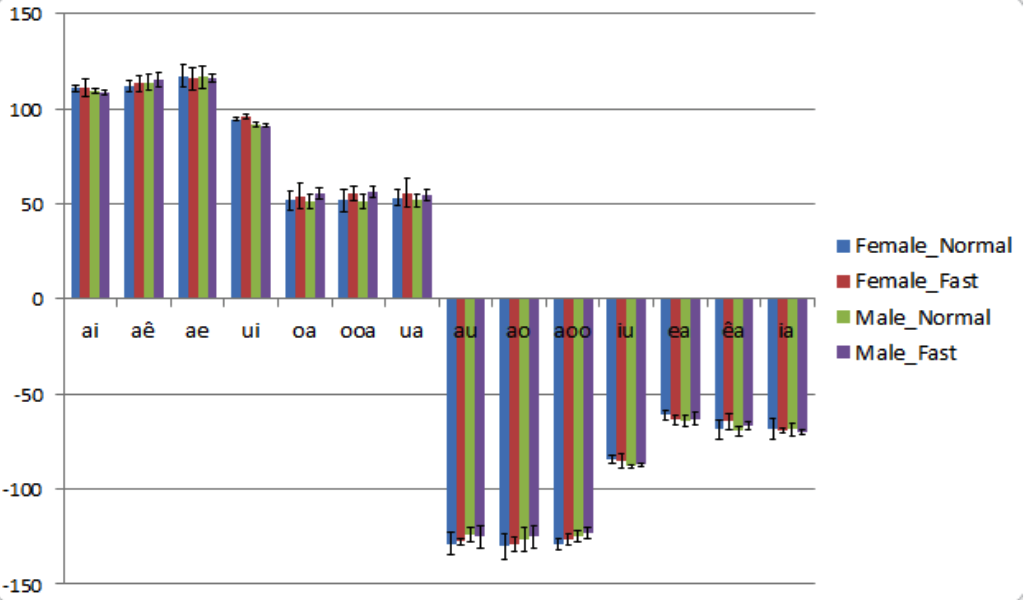}}
\caption{The SSCF angles on the SSCF1-SSCF2 plane at fast and normal rates produced by a Vietnamese male and female (source \protect\cite{tran2016acoustic}).}
\label{fig_sscf_angle_result}
\end{figure}

The VV trajectories can be considered straight lines on the SSCF1-SSCF2 plane (which corresponds roughly to the F1-F2 plane) (see Figure \ref{fig_ai_transition}). For each of these lines, it is then possible to define an angle with reference to the SSCF1 axis. For each pair of the transitions, V1V2 and V2V1, the sum of the two angles is equal to 180$^{\circ}$ (see Figure \ref{fig_sscf_angle}). Starting from this observation, the angles were proposed to characterize the direction of the vocalic transition between the adjacency frames on the SSCF planes. The angles were measured by hand over 80\% of the transition duration. According to the study, the angles were asserted to be speaker-independent because they were roughly the same across diverse speakers, including men and women, and similar results were obtained for the different speaking rates (see Figure \ref{fig_sscf_angle_result}) \cite{tran2016acoustic}.

\section{Proposed Method}
Previous work assumed that the spectral trajectories of vowel-to-vowel transitions are quasi-straight lines on SSCF1-SSCF2 plane. Thus it is possible to compute the angles to define the vocalic transitions \cite{tran2016acoustic}. The angles are generalized to calculate across all planes of $SSCF_i$-$SSCF_{i+1}$.

In each $SSCF_i$-$SSCF_{i+1}$ plane, the angle between a transition window of N frames are defined as in Equation \ref{formula_angle}.

\begin{equation}
\label{formula_angle}
Angle(j)= \frac{\pi}{180^\circ} \arctan(\frac{\Delta SSCF_{i+1}}{\Delta SSCF_i})
\end{equation}

Where j is a frame at index j; $\Delta SSCF_{i}$ is the difference between SSCF on the $i$th axis at the end and the beginning of the transition.

\begin{figure}[htbp]
\begin{center}
\includegraphics[width=7cm]{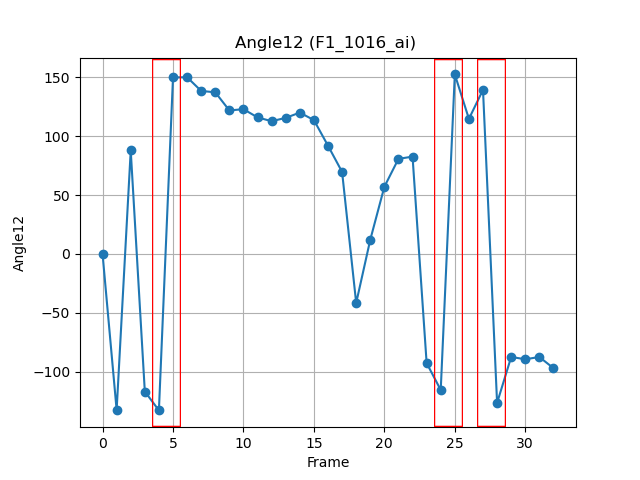}
\caption{The transition angles of /ai/ on the SSCF1-SSCF2 plane produced by a Vietnamese female speaker. The steps caused by the $\arctan$ function are highlighted in red rectangular.}
\label{fig_ai_angles}
\end{center}
\end{figure}

The transition angles are computed directly from the $\arctan$ function, which has a discontinuity at $-\pi$ and $+\pi$, resulting in the steps throughout the transition (see Figure \ref{fig_ai_angles}). Therefore, obtaining the angle directly from the $\arctan$ function is not a continuous parameter. Our experiments showed that using only $\arctan$ angles to define the acoustic trajectory of the speech signal is probably not a very good idea because the jumps will produce a kind of \quotes{noise} in the data.

That is why we propose to characterize the spectral trajectory in the acoustic space of the SSCFs using the polar coordinates. We assert that they give a more accurate representation because they are continuous variables and can specify not only the transition direction but also the trajectory of the spectrum. For a given frame, the two polar parameters (angle and radius) on the $SSCFi$-$SSCF_{i+1}$ plane are defined as in Equation \ref{formula_polar}.

\begin{equation}
\label{formula_polar}
\begin{aligned}
  Angle(j) &=  \frac{180^\circ}{\pi}\arctan(\frac{SSCF_{i+1}}{SSCF_{i}})\\
  Radius(j) &= \sqrt{SSCF_{i+1}^2 + SSCF_{i}^2}
\end{aligned}
\end{equation}

Where j represents a frame at index j; $SSCF_i$ and $SSCF_{i+1}$ represent the axis of the SSCF plane, respectively.

\section{Experiments}
The study made use of the Kaldi toolkit \cite{povey2011kaldi}. Note that the language model was trained using SRILM \cite{srilm} on the transcripts of the speech corpus because we want to measure the effect of the acoustic model, not the effect of out-of-vocabulary (OOV) words that existed in the test sets. The lexicon was constructed from the vocabulary of the language model using Phonetisaurus \cite{phonetisaurus}.

\subsection{Datasets}
The BRAF100 corpus \cite{vaufreydaz2000new} was used for French speech recognition. It contains around 28 hours of recordings from 100 native speakers (50 are men) aged from 15 to 63. Each speaker recorded between 102 and 105 different sentences and a common extract of \quotes{La Science et l'Hypothèse} of Henry Poincarée. The specification of the corpus is given in Table \ref{tbl_corpus}.

The corpus was prepared into three sets, each containing the train and test sets with comparable speaker sizes. The first set is for normal speech recognition, and the ASR model was trained and evaluated on the dataset of male and female speakers. The remaining two sets are for cross-gender speech recognition. The ASR model is trained on a male dataset and evaluated on a female dataset in cross-male recognition, whereas cross-female recognition does the opposite. The specification of datasets are specified in the Table \ref{tbl_french_datasets}.

{\renewcommand{\arraystretch}{1.35}%
\begin{table}
\caption{Specification of BRAF100}
\centering
\begin{tabular}{lcccc}
\hline
\hline
\bf Corpus & \bf Hour & \bf Transcript & \bf Speaker & \bf Vocab \\ 
\hline BRAF100 & 28h & 10,564 & 100 & 22,199  \\
\hline
\hline
\end{tabular}
\label{tbl_corpus}
\end{table}}

{\renewcommand{\arraystretch}{1.35}%
\begin{table*}[htbp]
\caption{The experimental datasets constructed from BRAF100}
\begin{center}
\begin{tabular}{ccccccc}
\hline
\hline
\multirow{2}{*}{\bf Experiment} & \multicolumn{3}{c}{\bf Train Set}                                                     & \multicolumn{3}{c}{\bf Test Set}                                                   \\ \cline{2-7} 
                            & \multicolumn{1}{c}{\bf Hour} & \multicolumn{1}{c}{\bf Transcript} & \bf Speaker              & \multicolumn{1}{c}{\bf Hour} & \multicolumn{1}{c}{\bf Transcript} & \bf Speaker           \\ \hline
Normal recognition          & \multicolumn{1}{c}{12h}  & \multicolumn{1}{c}{4,657}      & 22 males, 22 females & \multicolumn{1}{c}{3h}   & \multicolumn{1}{c}{1,036}      & 5 males, 5 females \\ \hline
Cross-male recognition      & \multicolumn{1}{c}{12h}  & \multicolumn{1}{c}{4,661}      & 44 males             & \multicolumn{1}{c}{1.5h} & \multicolumn{1}{c}{519}        & 5 females         \\ \hline
Cross-female recognition    & \multicolumn{1}{c}{13h}  & \multicolumn{1}{c}{5,278}      & 44 females           & \multicolumn{1}{c}{1.5h} & \multicolumn{1}{c}{517}        & 5 males           \\
\hline
\hline
\end{tabular}
\label{tbl_french_datasets}
\end{center}
\end{table*}}

\subsection{DNN-HMM Model}
\begin{figure}[ht]
\begin{center}
\includegraphics[width=7cm]{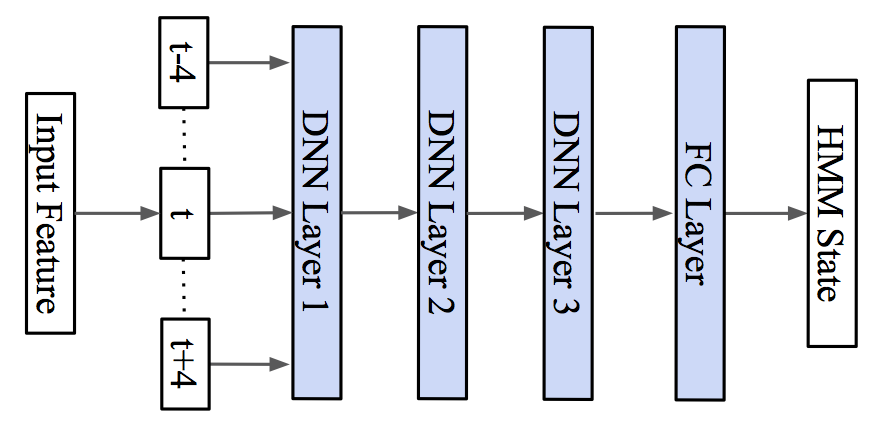}
\caption{Overview architecture of the DNN}
\label{fig_dnn}
\end{center}
\end{figure}

The ASR systems were constructed using a hybrid DNN-HMM model. The overall architecture of the DNN is given in Figure \ref{fig_dnn}. The network takes nine consecutive contextual frames as the input and feeds them into three hidden layers, each consisting of 512 neurons, and the hyperbolic tangent function (tanh) was employed as an activation function. These hidden layers are used to learn the representation of the spliced input features. The output from the hidden layers was then fetched into a fully connected (FC) layer to estimate the state of the triphone HMM using the softmax function.

Three training steps were conducted to produce the DNN-HMM model. First, a context-independent model (Monophone) was trained on the input features using 1K Gaussians over 40 iterations. Then, context-dependent (Triphone) training was performed over 35 iterations, utilizing 2K states and 10K Gaussians. Finally, the triphone DNN training was conducted across 20 epochs on the aligned training data from the context-dependent model. The initial and final learning rates of 0.01 and 0.001 were used during the training.

\subsection{Feature Configuration}
In this study, the angles and the polar coordinates were evaluated for characterizing the spectral trajectory of the speech signal on the SSCF planes and compared to the classical MFCC, the most widely used acoustic feature for speech recognition.

The angles were defined over a window size of one. The SSCFs were computed from short-time frames of 25 ms length with a 10 ms overlap, and the moving average was applied across a three-frame window to reduce the influence of the noise. A pre-emphasis factor of 0.97 and a Hamming window were utilized during feature extraction.

The MFCC with 6 and 13 dimensions were used as the baseline features. They were computed using 6 and 13 subband filters on the short-time frames with the same configuration as SSCFs, and a cepstral liftering coefficient of 22 was used while extracting the features.

\section{Results and Discussions}
The results in word error rate (WER) of normal and cross-gender speech recognitions are presented in Table \ref{tbl_result_braf100}. In normal recognition cases, the MFCC significantly outperformed the angle and polar coordinate. As expected, the 13 MFCC produced a greater result than the 6 MFCC because it captured more signal information (higher precision in the frequency domain). In addition, the performance was marginally improved when speed and acceleration were used ($\Delta, \Delta\Delta$).

The angle had the greatest error rate of 94\%, demonstrating that it is not a good parameter due to its discontinuity. However, the estimation of transition trajectory using polar coordinate surpassed angle by a wide margin. Furthermore, the performance was significantly improved when it was combined with speed and acceleration, particularly in cross-female recognition.

In the transgender recognition cases, similar patterns were obtained if the different types of parameters were compared. Nevertheless, all parameters had significantly higher error rates in all cases. It can be seen that the recognition rates for one gender are poorer if the learning is performed with speakers of the other gender. This shows that these parameters are not really independent of gender. 

In cross-male recognition, the 13 MFCC performed worse than the 6 MFCC, and the situation was quite similar when used with delta features ($\Delta, \Delta\Delta$). However, this was not the case for cross-female recognition. 

Overall, we still find that recognition rates are similar between the polar parameters and the MFCC (particularly for the 6 MFCC) compared to the angles.

However, the polar parameters characterizing the spectral trajectories do not seem much more gender-independent than the classical MFCC. This may be due to several aspects that we have not yet investigated in this first study: 

1) We considered SSCF0 (roughly corresponding to the fundamental frequency F0) participating in the same way as the other SSCFi similar to the formants. However, it is well known that the fundamental frequency F0 is quite different between men and women. Moreover, calculating a trajectory angle between SSCF0 and SSCF1 (F0 and F1) is probably not judicious because it does not correspond to real physical phenomena that SSCF0 (F0) characterizes the vibration of vocal cords, whereas the other SSCF characterize the resonances of the vocal tract.

2) A second point to investigate concerns the fact that characterizing the trajectories only by angles. If this is valid in the SSCF1-SSCF2 (F1-F2) plane where the trajectory during vowel-vowel transitions is indeed a quasi-straight line, it is not the case for the other $SSCFi$-$SSCF_{i+1}$ planes for $i>2$. We must consider characterizing this trajectory by another means.

3) A dynamic gesture is classically characterized by its trajectory and speed (and acceleration) of the movement on the trajectory. It is therefore important to characterize velocity and acceleration more judiciously than simply computing the first and second derivatives ($\Delta, \Delta\Delta$).

{\renewcommand{\arraystretch}{1.45}%
\begin{table*}[!htb]
\caption{Experimental results on BRAF100 in WER(\%)}
\begin{center}
\begin{tabular}{cccccc} 
  \hline
  \hline
    \thead{Experiment} & \thead{Parameter} & \thead{Dimension} & \thead{MONO} & \thead{TRI} & \thead{DNN} \\
  \hline
    \multirow{7}{11em}{\makecell{Normal recognition}} & 6 MFCC & 6 & \makecell{62.00} & \makecell{31.24} & \makecell{10.90} \\ 
    \cline{2-6} & 6 MFCC, $\Delta$, $\Delta \Delta$ & 18 & \makecell{26.70} & \makecell{11.20} & \makecell{08.65} \\ 
    \cline{2-6} & 13 MFCC & 13 & \makecell{41.90} & \makecell{17.09} & \makecell{08.39} \\ 
    \cline{2-6} & 13 MFCC, $\Delta$, $\Delta \Delta$ & 39 & \makecell{20.83} & \makecell{09.10} & \makecell{07.14} \\
    \cline{2-6} & Angle & 5 & \makecell{97.12} & \makecell{94.02} & \makecell{94.19} \\
    \cline{2-6} & Polar & 10 & \makecell{71.62} & \makecell{42.35} & \makecell{16.44} \\
    \cline{2-6} & Polar, $\Delta$, $\Delta \Delta$ & 30 & \makecell{46.17} & \makecell{21.10} & \makecell{13.01} \\
  \hline
    \multirow{7}{11em}{\makecell{Cross-male recognition}} & 6 MFCC & 6 & \makecell{67.41} & \makecell{39.58} & \makecell{16.51} \\ 
    \cline{2-6} & 6 MFCC, $\Delta$, $\Delta \Delta$ & 18 & \makecell{32.19} & \makecell{14.84} & \makecell{11.51} \\ 
    \cline{2-6} & 13 MFCC & 13 & \makecell{62.65} & \makecell{41.04} & \makecell{20.58} \\ 
    \cline{2-6} & 13 MFCC, $\Delta$, $\Delta \Delta$ & 39 & \makecell{36.39} & \makecell{23.12} & \makecell{16.74} \\
    \cline{2-6} & Angle & 5 & \makecell{96.87} & \makecell{97.29} & \makecell{94.60} \\
    \cline{2-6} & Polar & 10 & \makecell{79.25} & \makecell{56.81} & \makecell{28.43} \\
    \cline{2-6} & Polar, $\Delta$, $\Delta \Delta$  & 30 & \makecell{57.85} & \makecell{33.86} & \makecell{21.51} \\
  \hline
    \multirow{7}{11em}{\makecell{Cross-female recognition}} & 6 MFCC & 6 & \makecell{73.71} & \makecell{51.54} & \makecell{20.44} \\
    \cline{2-6} & 6 MFCC, $\Delta$, $\Delta \Delta$ & 18 & \makecell{35.63} & \makecell{17.59} & \makecell{15.30} \\ 
    \cline{2-6} & 13 MFCC & 13 & \makecell{66.96} & \makecell{43.92} & \makecell{18.62} \\
    \cline{2-6} & 13 MFCC, $\Delta$, $\Delta \Delta$ & 39 & \makecell{32.55} & \makecell{16.49} & \makecell{13.45} \\
    \cline{2-6} & Angle & 5 & \makecell{93.78} & \makecell{94.15} & \makecell{93.51} \\
    \cline{2-6} & Polar & 10 & \makecell{85.73} & \makecell{74.24} & \makecell{41.09} \\
    \cline{2-6} & Polar, $\Delta$, $\Delta \Delta$ & 30 & \makecell{64.31} & \makecell{48.24} & \makecell{29.20} \\
  \hline
  \hline
\end{tabular}
\label{tbl_result_braf100}
\end{center}
\end{table*}}

\section{Conclusion}
In this work, we investigated the use of polar coordinates in SSCF planes to describe the speech signal based on its acoustic trajectory. The finding showed that they significantly outperformed angles in both normal and cross-gender recognitions, demonstrating that the polar representation provides better information in characterizing the acoustic elements of the speech signal.

Even though the polar coordinates produced more significant results, there is still room for improvement. Our experiments showed that they are not significantly more gender-independent than the classical MFCC. We propose several directions for investigation in order to better account for the dynamic aspects of speech production. Examining how the cochlear system works will provide us with more useful ideas for characterizing the time aspect. Furthermore, it is desirable to investigate how to efficiently manage SSCF0 because using it directly with higher SSCFs is irrelevant. Finally, the SSCF0 must be normalized to account for spectral variation between speakers, particularly between male and female voices.

\subsection*{Acknowledgment}
This work was carried out as part of a PhD thesis funded by the French Embassy in Cambodia and the Cambodia Academy of Digital Technology.

\bibliographystyle{IEEEbib}
\bibliography{biblio.bib}

\end{document}